\documentstyle[aps,prl,multicol,epsf]{revtex}

\begin{document}
\title{The Density Matrix Renormalization Group in  Nuclear Physics: A Status Report}
\author{S.~Pittel$^1$, J. Dukelsky$^2$, S. Dimitrova$^3$ and M. Stoitsov$^{3-6}$}
\vspace{0.1in}
\address{
$^1$ Bartol Research Institute, University of Delaware, Newark, DE 19716 USA \\
$^2$ Instituto de Estructura de la Materia, CSIC, Serrano 123, 28006 Madrid, Spain \\
$^3$ Institute for Nuclear Research and Nuclear Energy, Bulgarian
Academy of Sciences, Sofia-1784, Bulgaria, \\
$^4$Physics Division, Oak Ridge National Laboratory, P.O. Box
2008, Oak Ridge, Tennessee 37831, USA\\
$^5$ Department of Physics \& Astronomy, University of Tennessee,
Knoxville, Tennessee 37996, USA, \\
$^6$Joint Institute for Heavy-Ion Research, Oak Ridge, Tennessee
37831, USA
 }
 \maketitle

\begin{abstract}
{We report on the current status of recent efforts to develop the
Density Matrix Renormalization Group method for use in large-scale
nuclear shell-model calculations.}
\end{abstract}

\begin{center}
\end{center}
\begin{multicols}{2}

\section{Introduction}

The Density Matrix Renormalization Group or DMRG  was originally
introduced by Steven White in the early 1990s to treat the
properties of quantum lattices\cite{White}. It quickly proved to
be enormously successful, producing for the ground state energy of
the spin-1 Heisenberg chain results that were accurate to 12
significant figures, well beyond what was achievable with any
other approximate many-body method. Subsequently, the method was
applied with great success to other 1D lattices, including spin
chains and t-J and Hubbard models\cite{book}. The model in its
original formalism, which worked in terms of real space lattice
sites, has also been applied, though with much less success, to
some 2D lattices as well.

Subsequently, the DMRG method was reformulated\cite{Xiang,WM} so
as not to work solely in terms of real space lattice sites. In
this extended version, the method has proven extremely useful in
the description of finite Fermi systems, as arise for example in
quantum chemistry\cite{QC} and in the physics of small metallic
grains\cite{sie}. The successes achieved in these domains suggests
the possible usefulness of the DMRG method in the description of
another finite Fermi system, the atomic nucleus. In this paper, we
briefly review the current status of our recent efforts to
implement such a program\cite{phDMRG,taxco01}.

The outline of the presentation is as follows. In Section II, we
briefly review the key elements of the DMRG method, whether for
quantum lattices or finite fermi systems. In Section III, we
describe our first efforts to apply a variant of the method to
real nuclei. As we will see, the results are not particularly
good. In Section IV, we describe a possible way to improve the
DMRG methodology, to more appropriately tune it to the the physics
of nuclei. Finally, in Section V, we summarize the key points of
the presentation.

\section{Brief Review of the DMRG formalism}

The DMRG is a method for systematically taking into account in an
approximate fashion all the degrees of freedom of a problem,
without letting the problem get numerically out of hand. The
method is rooted in Ken Wilson's ``onion" picture\cite{Wilson},
schematically illustrated in figure 1. Each layer of the onion
should be thought of as another degree of freedom. It could be a
site on a lattice or it could be a single-particle orbital in a
problem involving a finite fermi system.

\vspace*{0.5cm}
\begin{figure}
\hspace{1.5cm}  \epsfysize=6cm \epsfxsize=7cm \epsffile{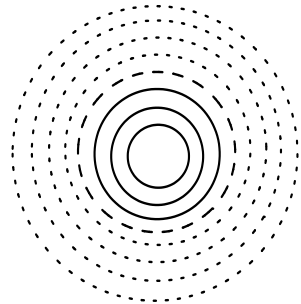}
\vspace{-0.8in} \narrowtext \caption{Schematic illustration of
Wilson's onion picture of the Renormalization Group.  The solid
lines refer to the layers already treated, i.e. comprising the
block; the dashed line refers to the layer added to enlarge the
block; the dotted lines refer to the layers still to be treated.}
\label{fig1}
\end{figure}

Assume that we have already taken into account a given number of
layers of the onion (illustrated by solid lines in the figure) and
that the total number of states we have kept to describe that
portion of the system, which we call the block, is $p$.  Assume
furthermore that the next layer of the onion (illustrated by a
dashed line) contains another $s$ states. If we add the new layer
to the block, the total number of states in the {\em enlarged
block} will be $p \times s$. The basic idea of all Renormalization
Group Methods, whether Wilson's numerical Renormalization Group
(RG) or White's DMRG, is to somehow truncate from these $p \times
s$ states to the most important $p$ states, the same number we had
before enlarging the block. We would then add the next layer of
the onion, truncating afterwards to the most important $p$ states
again. And we would do this until all layers have been treated. We
would carry out the calculation as a function of $p$, stopping
when the change with increasing $p$ is sufficiently small.

To this point, we have not yet said how to implement the
truncation. Indeed, the answer is different in the usual Wilson RG
method and in White's DMRG method.

In Wilson's method, one simply diagonalizes the hamiltonian in the
enlarged block, including the extra layer, and then truncates to
the lowest $p$ eigenstates.

In White's DMRG method the idea is different, and better. Here,
one considers the enlarged block in a medium that approximates the
rest of the system. The entire system - enlarged block + medium -
is referred to as the superblock. The hamiltonian is diagonalized
in the superblock, yielding a ground state wave function

\begin{equation}
|\Psi> ~=~ \sum_{i=1,p \times s} \sum_{j=1,t} \Psi_{ij} |i>_B
|j>_M
\end{equation}
where $B$ denotes the states in the block, $M$ the states in the
medium and $t$ the number of states in the medium.

The reduced density matrix for the enlarged block in the ground
state,

\begin{equation}
\rho^B_{ii'}= \sum_{j=1,t} \Psi_{ij}  \Psi^{*}_{i'j} ~,
\end{equation}
is then constructed and diagonalized,
\begin{equation}
\rho^B | u^{\alpha} >_B ~=~ \omega^B_{\alpha} ~ |u^{\alpha} >_B ~.
\end{equation}

Those eigenstates $|u^{\alpha}>_B$ with the largest eigenvalues
$\omega^B_{\alpha}$ are the most important states of the enlarged
block in the ground state of the superblock, i.e. of the system.
One thus truncates to the $p$ states of the enlarged block with
the largest density matrix eigenvalues.

Details on how this can be efficiently done can be found in the
Proceedings of this series of symposia in 2001 \cite{taxco01}.

\subsection{The finite versus infinite algorithm}

The method described above, where one passes through the set of
onion layers a single time, is referred to as the infinite DMRG
algorithm. Depending on the manner in which correlations between
layers fall off, this method can sometimes lead to an accurate
representation of the ground state of the system and perhaps some
excited states as well . Usually, it does not, however, since the
early layers know nothing of the physics of those treated
subsequently. Thus the blocks are only being optimized with
respect to the layers that make them up. This suggests the use of
a ``sweeping" algorithm, whereby once all layers have been sampled
we reverse direction and update the blocks based on the
information of the previous ``sweep". Such a sweeping algorithm
can be iteratively implemented until acceptable convergence in the
results has been achieved. The latter procedure, in which we sweep
back and forth through the onion, is called the finite algorithm.
This is the method usually needed when dealing with finite fermi
systems, like nuclei.

\subsection{The p-h DMRG method}

In the description of finite fermi systems, it is natural to use a
basis of single-particle states and to implement the DMRG by
iteratively adding their effects. An important feature of such
systems is the presence of a fermi surface, which divides the set
of single-particle levels into those that are primarily occupied
and those that are primarily filled. It is possible to incorporate
this into the DMRG algorithm in the following way.

\begin{figure}
\hspace{-0.5cm} \epsfysize=8cm \epsfxsize=8cm\epsffile{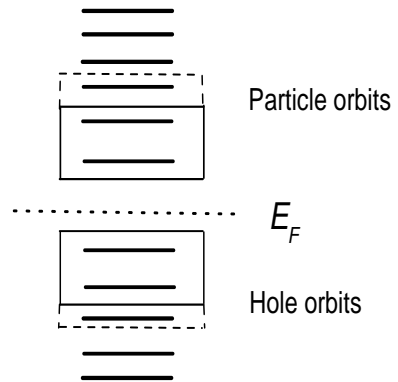}
\vspace{-1in} \narrowtext \caption{Schematic illustration of the
p-h DMRG method for finite fermi systems.  The single-particle
levels are divided into primarily occupied hole levels and
primarily empty particle levels. The first two levels above the
fermi energy $E_F$ have been formed into a particle block and the
first two below the fermi energy have been formed into a hole
block. The third levels above and below are then added to form
enlarged blocks.} \label{fig2}
\end{figure}

Consider the set of single-particle levels shown in figure 2. Note
that they are divided into two sets. There are particle levels -
those that are primarily empty - and hole levels - those that are
primarily filled. In the p-h DMRG method, one starts with the
levels nearest to the fermi surface, which make up particle and
hole blocks, and then gradually adds to them levels further away.
The figure is meant to signify that we have already created
particle and hole blocks involving the first two available levels,
respectively, and that we then enlarge them by adding the third
level(s). The particle block then serves as the medium for the
holes and the hole block as the medium for the particles as we
implement the DMRG truncation strategy.  This process is continued
until all particle and all hole levels have been treated.

The p-h DMRG method is particularly useful when the infinite
algorithm can be used. When it is necessary to introduce sweeping,
it becomes too cumbersome to implement because of the
preponderance of blocks that have to be considered. This is
especially true for nuclei in which one has two kinds of particles
and thus twice as many blocks. In such cases, there are four
distinct blocks -- proton particle, proton hole, neutron particle
and neutron hole -- and this makes it very hard to implement
sweeping.

Nevertheless, the first calculations carried out for nuclei using
the DMRG method were based on the p-h algorithm. The hope was that
the correlations would fall off sufficiently rapidly as we
progress away from the fermi surface, making the infinite
algorithm acceptable. It is these results, the first that we have
obtained using the DMRG for realistic nuclei, that will be
presented in the next section.

\section{Application of the p-h DMRG method to $^{24}Mg$}

As a first application of the p-h DMRG method in nuclear
structure, we considered the nucleus $^{24}Mg$, with four neutrons
and four protons outside doubly-magic $^{16}O$. As is usual in the
shell model, we assumed that $^{16}O$ is inert and distributed the
remaining 8 nucleons over the orbits of the $2s-1d$ shell only.
This shell-model problem is small -- perhaps too small -- and can
be solved trivially by exact shell-model diagonalization.

\begin{figure}
\hspace{-0.5cm} \epsfysize=8cm \epsfxsize=8cm\epsffile{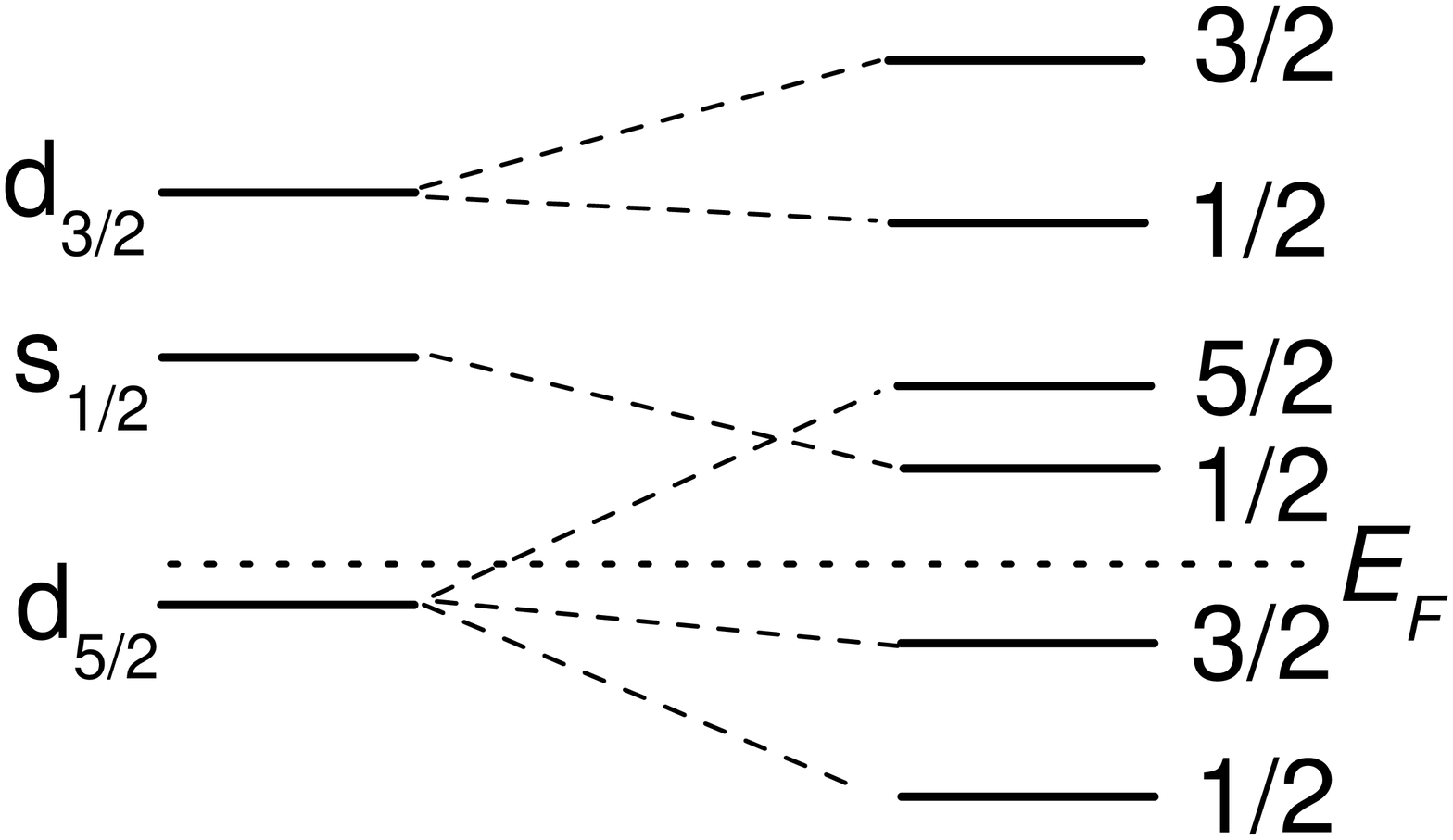}
\vspace{-1in} \narrowtext \caption{Schematic illustration of the
splitting of the spherical single--particle levels of the
$sd$--shell into a set of doubly--degenerate levels by an
axially--deformed Hartree--Fock calculation. The dashed line
represents the Fermi energy ($E_F$), which separates the particle
levels from the hole levels. Each doubly--degenerate level is
labelled by its angular momentum projection on the intrinsic
z-axis.} \label{fig3}
\end{figure}

The results we show were obtained using a Hartree Fock
single-particle basis to define the single-particle levels and
their order. These levels are illustrated in figure 3.

There is a great calculational simplification when one uses a
single-particle basis in which all levels are essentially the
same. And indeed when one uses an axially deformed HF basis, all
levels are doubly degenerate and thus very similar in structure.

The calculations were done using the usual USD effective
hamiltonian for the $2s-1d$ shell.

In figure 4, we show results for the energies of the lowest four
states of the nucleus.  The results are presented as a function of
the quantity $p$ introduced earlier, namely the number of states
kept in a block. For this problem the largest $p$ that can be
achieved is $p=64$. The solid line in each panel gives the exact
result.

\begin{figure}
\vspace{-0.5cm} \hspace*{-1cm} \epsfysize=14cm
\epsfxsize=14cm\epsffile{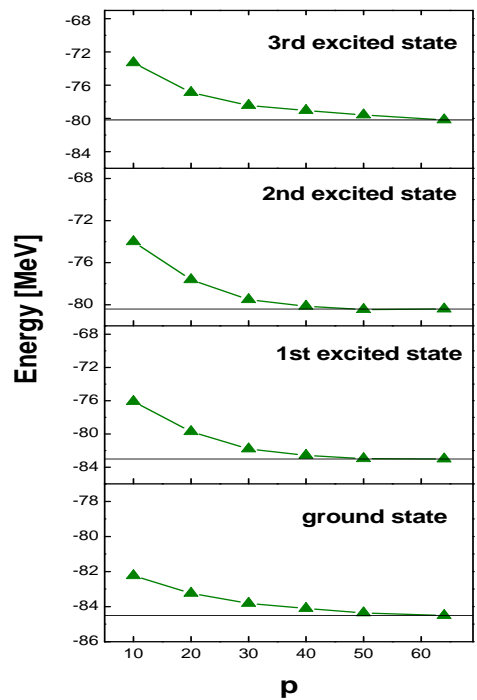}
\vspace*{-3cm} \narrowtext \caption{The energy of the ground state
and of the three lowest excited states for $^{24}Mg$ as a function
of $p$, the number of states retained in the block. The horizontal
solid lines refer to the exact eigenenergy for each state. }
\label{fig4}
\end{figure}

Two points should be noted. First, the results indeed converge to
the exact results for $p=64$, as they should. What is more
important to note, however, is that the convergence is very slow.
We need to have values of $p$ above $40$ to get reasonably
accurate results when compared with the exact values. For the
values of $p$ needed to get an accurate reflection of the full
results we must treat hamiltonian matrices not much smaller than
those of the full problem. Clearly, the method -- as just
described -- is not working very well.

\subsection{What's wrong?}

From our perspective, there are at least two things wrong with the
calculational framework that was used. On the one hand, since we
worked in the m-scheme, we did not preserve angular momentum when
we imposed truncation. If we want to work in the m-scheme and not
lose angular momentum conservation we would have to (1) work in
terms of spherical single-particle states, (2) include all states
from a given orbit in a single step, and (3) make sure never to
cut the truncation within a set of degenerate density matrix
eigenvalues. This is possible, but extremely difficult, especially
for systems with both neutrons and protons.

The second limitation of the method we employed is that it used
the infinite algorithm and thus did not implement sweeping.  As
noted earlier, it is not practical to include sweeping in the p-h
algorithm because of the preponderance of blocks that need to be
coupled together. We have indeed implemented sweeping for
$^{24}Mg$, and it did improve the results slightly, but it is
clear that this is not the way to proceed for more complex nuclei.

\section{What next?}

Based on these considerations, we now believe that the minimum
requirements for a useful DMRG strategy in nuclear physics are (1)
that it works in a J-scheme or angular momentum conserving basis
and (2) that it implements sweeping. In the following subsection,
we sketch how this can be done. Further discussion of
symmetry-conserving DMRG methods can be found in ref.
\cite{Gyulacsi}.

\subsection{The J-DMRG}

\subsubsection{Initialization}

We will illustrate the J-DMRG method through a problem involving
five neutron orbits and five proton orbits, as illustrated in
figure 5. These are spherical shell-model orbits, with definite
$n$, $l$ and $j$. However, we only show $j$ for simplicity.

\large

\begin{center}
$
\begin{tabular}{lllll}
\multicolumn{1}{l}{$j_{5\quad }\_\_\_\_\quad $} &
\multicolumn{1}{l}{} &
&  & \multicolumn{1}{l}{$\quad \_\_\_\_\quad j_{10}$} \\
\multicolumn{1}{l}{$j_{4\quad }\_\_\_\_\quad $} &
\multicolumn{1}{l}{} & &  & \multicolumn{1}{l}{$\quad
\_\_\_\_\quad j_{9}$} \\
\multicolumn{1}{l}{$j_{3\quad }\_\_\_\_\quad $} &
\multicolumn{1}{l}{} & &  & $\quad \_\_\_\_\quad j_{8}$ \\
\multicolumn{1}{l}{$j_{2}\quad \_\_\_\_\quad $} &
\multicolumn{1}{l}{} &
&  & \multicolumn{1}{l}{$\quad \_\_\_\_\quad j_{7}$} \\
\multicolumn{1}{l}{$j_{1\quad }\_\_\_\_\quad $} &
\multicolumn{1}{l}{} &
&  & \multicolumn{1}{l}{$\quad \_\_\_\_\quad j_{6}$} \\
 \multicolumn{1}{c}{$\nu $} &  &  &  &
\multicolumn{1}{c}{$\pi $}
\end{tabular}
$
\end{center}

\normalsize
\begin{figure}
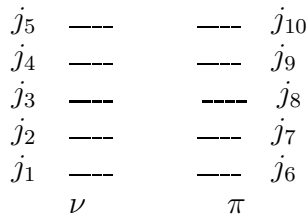

 \narrowtext \caption{Active single-particle levels
for the discussion of the J-DMRG method in the text.} \label{fig5}
\end{figure}

Input to the calculation includes the number of neutrons and
protons, the shell-model hamiltonian, including one- and two-body
terms, and a set of single-shell matrix elements for each active
orbit.

The single-shell matrix elements are the reduced matrix elements
of all sub-operators of the hamiltonian within the orbit.  All can
be readily obtained from the Coefficients of Fractional Parentage
for the orbit. Such single-shell reduced matrix elements are a
common feature of all J-scheme shell model codes.

\subsubsection{The warm-up phase}

The next step is the warm-up phase, in which we carry out a first
pass through the orbits and store an initial set of reduced matrix
elements associated with the various possible blocks (groups of
orbits). This includes for neutrons the orbits $j_1 \rightarrow
j_2$, $j_1 \rightarrow j_3$, $j_1 \rightarrow j_4$ and $j_1
\rightarrow j_5$, with corresponding blocks for protons.

\large

\begin{center}
$
\begin{tabular}{lllll}
\multicolumn{1}{l}{$j_{5\quad }\_\_\_\_\quad $} &
\multicolumn{1}{l}{} &
&  & \multicolumn{1}{l}{$\quad \_\_\_\_\quad j_{10}$} \\
\multicolumn{1}{l}{$j_{4\quad }\_\_\_\_\quad $} &
\multicolumn{1}{l}{} & &  & \multicolumn{1}{l}{$\quad
\_\_\_\_\quad j_{9}$} \\
\multicolumn{1}{l}{$j_{3\quad }\_\_\_\_\quad $} &
\multicolumn{1}{l}{} & &  & $\quad \_\_\_\_\quad j_{8}$ \\
\cline{1-1} \cline{5-5} \multicolumn{1}{|l}{$j_{2}\quad
\_\_\_\_\quad $} & \multicolumn{1}{|l}{} &
&  & \multicolumn{1}{|l|}{$\quad \_\_\_\_\quad j_{7}$} \\
\multicolumn{1}{|l}{$j_{1\quad }\_\_\_\_\quad $} &
\multicolumn{1}{|l}{} &
&  & \multicolumn{1}{|l|}{$\quad \_\_\_\_\quad j_{6}$} \\
\cline{1-1}\cline{5-5} \multicolumn{1}{c}{$\nu $} &  &  &  &
\multicolumn{1}{c}{$\pi $}
\end{tabular}
$
\end{center}

\normalsize
\begin{figure}
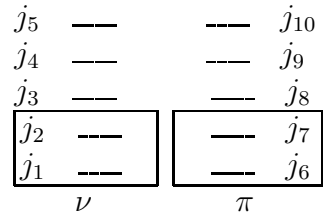

\narrowtext \caption{Schematic illustration of the warm-up phase
of the J-DMRG method for finite fermi systems.  The lowest two
neutron levels form a neutron block and the lowest two proton
levels a proton block. } \label{fig6}
\end{figure}

Consider for example the schematic illustration shown in figure 6.
At this point, we have treated the first two orbits for both
neutrons and protons. In each of the two blocks, we have a given
number of states with each $n$ (the total number of particles) and
$J$ (the total angular momentum). The number of states is
analogous to the quantity $p$ introduced earlier. Furthermore, in
the two blocks the reduced matrix elements of all sub-operators of
the hamiltonian have already been stored.

We then add the next neutron level, as illustrated in figure 7.

\large

\begin{center}
$
\begin{tabular}{lllll}
\multicolumn{1}{l}{$j_{5\quad }\_\_\_\_\quad $} &
\multicolumn{1}{l}{} &
&  & \multicolumn{1}{l}{$\quad \_\_\_\_\quad j_{10}$} \\
\multicolumn{1}{l}{$j_{4\quad }\_\_\_\_\quad $} &
\multicolumn{1}{l}{} & &  & \multicolumn{1}{l}{$\quad
\_\_\_\_\quad j_{9}$} \\
\cline{1-1} \multicolumn{1}{|l}{$j_{3\quad }\_\_\_\_\quad $} &
\multicolumn{1}{|l}{} & &  & $\quad \_\_\_\_\quad j_{8}$ \\
\cline{1-1} \cline{5-5} \multicolumn{1}{|l}{$j_{2}\quad
\_\_\_\_\quad $} & \multicolumn{1}{|l}{} &
&  & \multicolumn{1}{|l|}{$\quad \_\_\_\_\quad j_{7}$} \\
\multicolumn{1}{|l}{$j_{1\quad }\_\_\_\_\quad $} &
\multicolumn{1}{|l}{} &
&  & \multicolumn{1}{|l|}{$\quad \_\_\_\_\quad j_{6}$} \\
\cline{1-1}\cline{5-5} \multicolumn{1}{c}{$\nu $} &  &  &  &
\multicolumn{1}{c}{$\pi $}
\end{tabular}
$
\end{center}

\normalsize

\begin{figure}
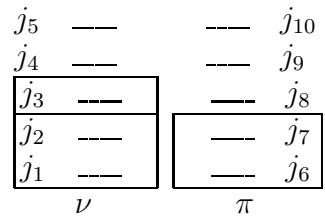

\narrowtext \caption{Schematic illustration of the enlargement of
the neutron block of figure 6 in the warm-up phase of the J-DMRG
method. } \label{fig7}
\end{figure}

To do so, we first construct states of good angular momentum for
the enlarged neutron block. We then calculate the matrix elements
of all neutron operators in the enlarged neutron block, using
standard formulae given for example in ref. \cite{deShalit}. The
calculation requires reduced matrix elements from the block made
up of orbits $j_1$ and $j_2$, which were stored in the previous
iteration, and those of the orbit $j_3$ , which were stored in the
input stage.

We then couple neutrons and protons together to states of good
angular momentum, calculate the hamiltonian matrix in this basis
and diagonalize. We then construct the ground state density matrix
for neutrons and use it to truncate to the same number of states
as we had before level $j_3$ was added. Then we transform all
reduced matrix elements to the truncated block and store them.

And then we add the next proton level, and then the next neutron
level, etc, until all levels have been included. At that point we
have a first guess as to the optimum states associated with the
successively larger blocks of levels.

\subsubsection{The sweeping phase}

We now turn to the sweeping phase, schematically illustrated in
figure 8. It is during this phase that we systematically improve
our description of the physics of each of the blocks, by taking
into account the influence of the other orbits in the problem.

The idea of the picture is as follows. We have just treated the
block of proton orbits $j_9$ and $j_{10}$. We now wish to add to
it the proton orbit $j_8$ to create an enlarged proton block. We
will carry out a truncation in this enlarged block, following the
density matrix strategy. To do so, we consider the block $j_6
\rightarrow j_7$ as the proton medium and the block $j_1
\rightarrow j_5$ as the neutron medium. We then couple the states
of the enlarged proton block to the two parts of the medium to
obtain the superblock, an approximation to the entire system. We
calculate the hamiltonian in the associated superblock, using only
stored information  We then diagonalize the superblock hamiltonian
and determine the density matrix for the enlarged proton block,
orbits $j_8 \rightarrow j_{10}$. We then use this to truncate the
enlarged proton block to the same number of states as we had
before enlarging it. Then, we calculate all reduced matrix
elements in the truncated proton block and store them. And then we
add the next orbit, $j_7$. We do this for all proton blocks and
then for all neutron blocks. And after the sweep is finished, we
simply turn around and sweep upwards, continuing the process until
the results from one sweep and those from the previous sweep are
acceptably close.

\vspace*{0.5cm}

\large

\begin{center}
$
\begin{tabular}{lllll}
\cline{1-1}\cline{5-5} \multicolumn{1}{|l}{$j_{5\quad
}\_\_\_\_\quad $} & \multicolumn{1}{|l}{} &
&  & \multicolumn{1}{|l|}{$\quad \_\_\_\_\quad j_{10}$} \\
\multicolumn{1}{|l}{$j_{4\quad }\_\_\_\_\quad $} &
\multicolumn{1}{|l}{} & &  & \multicolumn{1}{|l|}{$\quad
\_\_\_\_\quad j_{9}$} \\ \cline{5-5}
\multicolumn{1}{|l}{$j_{3\quad }\_\_\_\_\quad $} &
\multicolumn{1}{|l}{} & &  & $\quad \_\_\_\_\quad j_{8}$ \\
\cline{5-5} \multicolumn{1}{|l}{$j_{2}\quad \_\_\_\_\quad $} &
\multicolumn{1}{|l}{} &
&  & \multicolumn{1}{|l|}{$\quad \_\_\_\_\quad j_{7}$} \\
\multicolumn{1}{|l}{$j_{1\quad }\_\_\_\_\quad $} &
\multicolumn{1}{|l}{} &
&  & \multicolumn{1}{|l|}{$\quad \_\_\_\_\quad j_{6}$} \\
\cline{1-1}\cline{5-5} \multicolumn{1}{c}{$\nu $} &  &  &  &
\multicolumn{1}{c}{$\pi $}
\end{tabular}
$
\end{center}

\normalsize

\begin{figure}
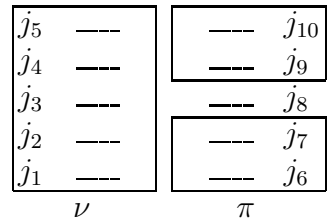

\narrowtext \caption{Schematic illustration of the sweeping phase
of the J-DMRG algorithm. We are sweeping downwards in protons,
having treated as a block the levels $j_9$ and $j_{10}$. The level
$j_8$ is then to be added to the block, and the enlarged block is
treated in a medium involving the proton block $j_6 \rightarrow
j_7$ and the neutron block $j_1 \rightarrow j_5$ from the previous
sweep. } \label{fig8}
\end{figure}

The formalism, as just described, is in the process of being
implemented by two of the authors (SP and JD) in collaboration
with Nicu Sandulescu and Larisa Pacearescu.

\section{Closing Remarks}

In this presentation, we have provided a status report on recent
efforts to develop the Density Matrix Renormalization Group method
for use in large-scale nuclear shell-model calculations. Following
a brief review of the general ideas behind the DMRG, we described
the first application of the p-h variant of the method to
realistic nuclei. Regrettably, the results were not especially
promising. We discussed some of the reasons for the failure and
then discussed a possible strategy that might overcome those
shortcomings. The basic idea is to implement the DMRG algorithm in
an angular-momentum-conserving basis and to include sweeping.
While we are guardedly optimistic that this will indeed provide a
practical and efficient methodology for large-scale nuclear
structure calculations of heavy nuclei, only time will tell.

\vspace{1cm}

{\bf Acknowledgments} This work was supported in part by the US
National Science Foundation under grant \#s PHY-9970749 and
PHY-0140036, by the Spanish DGI under grant BFM2000-1320-C02-02,
by NATO under grant PST.CLG.977000, and by the Bulgarian National
Foundation for Scientific Research under Contract \# $\Phi$-809.
One of the authors (SP) wishes to acknowledge fruitful discussions
on the J-DMRG method with Nicu Sandulescu.

\end{multicols}

\vspace{-2ex}

\begin{thebibliography}{9}
\bibitem{White}  S. R. White, Phys. Rev. Lett. {\bf 69}, 2863
(1992); S. R. White, Phys. Rev. {\bf B48}, 10345 (1993); S. R.
White and D. A. Huse, Phys. Rev. {\bf B48}, 3844 (1993).

\bibitem{book}  {\it Density Matrix Renormalization}, edited by I. Peschel,
X. Xiang, M. Kaulke and K. Hallberg, {\sl Lectures Notes in
Physics}, (Springer. Berlin, 1999); S. R. White, Phys. Rep. {\bf
301}, 187 (1998).

\bibitem{Xiang}
T. Xiang, Phys. Rev. {\bf B53} (1996) R10445.

\bibitem{WM}  S. R. White and R. L. Martin, J. Chem. Phys. {\bf 110}, 4127
(1999).

\bibitem{QC}  S. Daul, I. Ciofini, C. Daul and S. R. White, Int. J. Quantum
Chem. {\bf 79}, 331 (2000); A. O. Mitrushenko, G. Fano, F.
Ortolani, R. Linguerri and P. Palmeri, J. Chem. Phys. {\bf 115};
G. Fano, F. Ortolani and L. Ziosi, J. Chem. Phys. {\bf 108}, 9246
(1998); L. Bendazzoli, S. Evangelisti, G. Fano, F. Ortolani and L.
Ziosi, J. Chem. Phys. {\bf 109}, 1277 (1999) 6815 (2001); G. K-L.
Chan and M. Head-Gordon, J. Chem. Phys. {\bf 116} 4462 (2002).

\bibitem{sie}  J. Dukelsky and G. Sierra, Phys. Rev. Lett. {\bf 83}, 172
(1999); J. Dukelsky and G. Sierra, Phys. Rev. {\bf B61}, 12302
(2000).

\bibitem{phDMRG}  J. Dukelsky and S. Pittel, Phys. Rev. {\bf C63}, 061303(R)
(2001); J. Dukelsky, S. Pittel, S.S. Dimitrova, M.V. Stoitsov,
Phys. Rev. {\bf C65}, 054319 (2002).

\bibitem{taxco01} S. Pittel and J. Dukelsky, Rev. Mex. de F\'{i}sica {\bf 47
Suppl.} (2001) 42.


\bibitem{Wilson}  K. G. Wilson, Rev. Mod. Phys. {\bf 47}, 773 (1975).

\bibitem{Gyulacsi} Ian P. McCulloch and Mikl\'{o}s Gul\'{a}csi,  Europhys.
Lett. {\bf 57}, 852 (2002)

\bibitem{deShalit} A. de-Shalit and I. Talmi, {\sl Nuclear Shell
Theory}, (Academic Press, New York, 1963).

\end{thebibliography}
\end{document}